\documentclass[11pt,a4]{article}
\usepackage{asaf}

\title{Almost Tight Bounds for Tree Covers}

\author{
  Asaf Petruschka%
  \thanks{
  Supported by an Azrieli Foundation fellowship. 
  Email: \texttt{asaf.petruschka@weizmann.ac.il}
  } 
  \qquad 
  Nir Petruschka\thanks{
    Email: \texttt{nir.petruschka@weizmann.ac.il}
  } 
  \qquad
  Elad Tzalik\thanks{Supported by the Adams Fellowship Program of the Israel Academy of Sciences and Humanities.
  Email: \texttt{elad.tzalik@weizmann.ac.il}.}
\\  Weizmann Institute of Science
}

\date{}

\begin{document}
\maketitle

\begin{abstract}

We show that every $n$-point metric space admits a cover by $k$ tree metrics, such that the distance between each pair of points is approximated by their distance in one of the trees up to a factor of $n^{O(1/k^2)} \cdot \log n$. %
The previous best upper bound of $\tilde{O}(n^{1/k})$ was given by Bartal, Fandina and Neiman (JCSS 2022).
This closes the gap to the recent $\Omega_k (n^{1/(k-1)^2})$ lower bound of Chen, Huang, Tan, and Xu (2026) up to constant factors in the exponent.
The result is obtained via a simple reduction that transforms a given \emph{Ramsey} tree cover with $k$ trees into a non-Ramsey one with $O(\sqrt{k})$ trees, while preserving the approximation factor up to constants.

\end{abstract}

\paragraph{AI disclosure.}
The results were initially discovered using ChatGPT Astra 6.1-2. The authors have verified, simplified and rewritten the proofs. The exposition is due to the authors, who assume full responsibility for all content. 

\section{Introduction}

\emph{Tree covers} approximate a given metric space by tree metrics, which are often easier to work with for algorithmic applications.
Formally, let $(X,d_X)$ be an $n$-point metric space.

\begin{definition}
A \emph{dominating tree} is a weighted tree $T$ with an injective\footnote{The embedding need not be surjective, that is, $T$ may have additional \emph{Steiner} nodes that do not correspond to any point in $X$. By~\cite{Gupta01}, these Steiner nodes can be removed while increasing the distances between the (nodes corresponding to) points in $X$ in the tree $T$ by at most a factor of $8$. That is, if $x,y \in X$ had distortion $\alpha$ before removing Steiner nodes, then $x,y$ have distortion $8\alpha$ after removing them.} embedding $t: X \to V(T)$ such that $d_{T} (t(x),t(y)) \geq d_X (x,y)$ for all $x,y \in X$.
The tree $T$ has \emph{distortion} $\alpha$ for $x,y \in X$ if $d_T (t(x),t(y)) \leq \alpha \cdot d_X (x,y)$.
For notational convenience, we identify $t$ with $T$ and write $T(x) := t(x)$ and $d_T (x,y) := d_T (t(x),t(y))$ for $x,y \in X$.
\end{definition}

\begin{definition}
    A collection of dominating trees $T_1, \dots, T_k$ is an \emph{$\alpha$-tree cover} if for every pair $x,y \in X$, some $T_i$ has distortion $\alpha$ for $x,y$.
    It is an \emph{$\alpha$-Ramsey tree cover} if there is also a mapping of $X$ to \emph{home trees} $\home : X \to [1..k]$ such that, for every $x \in X$, $T_{\home (x)}$ has distortion $\alpha$ for $x,y$, for every other $y \in X$.    
\end{definition}

Using trees to approximate a finite metric space allows one to exploit existing techniques used for tree metrics in a general setting.
The usefulness of this approach is perhaps surprising, since some $n$-point metrics are ``far'' from being tree metrics; namely, embedding them into a tree metric requires distortion $\Omega(n)$~\cite{RR98}.
One approach to overcome this obstacle is by using probabilistic embeddings into tree metrics, that is, to approximate the metric using a distribution over trees instead of just a single tree.
This classical notion allows one to preserve the distance between every pair of points in expectation up to a factor of $O(\log n)$~\cite{Bartal96, Bartal98, CCGGP98, FRT04, Bartal04}.
Tree covers also overcome this obstacle, but with a deterministic guarantee that, for each pair of points, at least one tree in the cover approximately preserves their distance.
Due to their algorithmic applications to problems such as routing~\cite{CGMZ05, ACEFN18}, oblivious network design~\cite{GHR06}, and the construction of distance oracles~\cite{MN07}, (Ramsey) tree covers have been studied extensively in many natural settings (see, for example, ~\cite{CGMZ05, BFN22, DCCLST25, CCLMST26}).

The current state-of-the-art upper bound on tree covers for general $n$-point metrics is by Bartal, Fandina and Neiman~\cite{BFN22}, which in fact gives \emph{Ramsey} tree covers:
\begin{theorem}[\cite{BFN22}]\label{thm:bfn}
Every $n$-point metric has an $O\bigl(n^{1/k}(\log n)^{1-1/k}\bigr)$-Ramsey tree cover with $k$ trees, for all $k \geq 1$.
\end{theorem}
It was also shown in~\cite{BFN22} that this distortion is tight for Ramsey tree covers (up to the $(\log n)^{1-1/k}$ factor).
On the lower bound side, only recently Chen, Tan, and Xu~\cite{CTX26} gave the first polynomial
(in $n$) bound of%
\footnote{Throughout, the notation $\tilde{O}(f)$ hides $\poly(\log f)$ factors,
and $O_\alpha(\cdot)$ hides a factor that depends only on $\alpha$.}
$\Omega_k(n^{1/2^{k-1}})$ for the distortion of (non-Ramsey) tree covers with $k$ trees, which was subsequently improved to $\Omega_k(n^{1/(k-1)^2})$ for all $k \geq 3$ by Chen, Huang, Tan, and Xu~\cite{CHTX26}.

Our main contribution 
is a reduction from Ramsey tree covers with $k$ trees to regular tree covers with $O(\sqrt{k})$ trees, while only incurring a loss of factor $3$ in distortion.
Formally,
\begin{theorem}[Main Result]\label{thm:main}
    If an $n$-point metric space $(X, d_X)$ admits an $\alpha$-Ramsey tree cover with $k$ trees, then it also admits a $3\alpha$-tree cover with $O(\sqrt{k})$ trees.
\end{theorem}

The proof is given in Section~\ref{sec:proof-of-main}.
Applying this reduction to Theorem~\ref{thm:bfn} yields an upper bound for tree covers which matches the lower bound from~\cite{CHTX26} up to constant factors in the exponent:

\begin{corollary}
Every $n$-point metric space admits an $(n^{O(1/k^2)} \cdot \log n)$-tree cover with $k$ trees, for all $k \geq 1$. 
\end{corollary}

As a warm-up to~\Cref{thm:main}, we show in Section~\ref{sec:warmup} a reduction from an $\alpha$-Ramsey tree cover with $4$ trees to a (non-Ramsey) $\alpha$-tree cover with $3$ trees, i.e., the distortion remains exactly $\alpha$.
Combining with~\Cref{thm:bfn}, this gives near-optimal distortion for covers with $3$ trees:
\begin{corollary}\label{cor:3-cover}
    Every $n$-point metric admits an $O(n^{1/4} \log^{3/4} n)$-tree cover with $3$ trees.
\end{corollary}

From now on, we fix the $n$-point metric space $(X,d_X)$, and a given $\alpha$-Ramsey tree cover $T_1, \dots, T_k$ of $X$ with home assignment $\home : X \to [1..k]$.
We denote $H_i = \home^{-1}(i)$, that is, $H_i$ is the set of all $x \in X$ whose home tree is $T_i$.

\section{Warm-up: Tight Bounds for Covers with $3$ Trees}\label{sec:warmup}

Let us assume that the $\alpha$-Ramsey tree cover has $4$ trees $T_1, T_2, T_3, T_4$.
Our goal in this warm-up is to prove that we can use it to construct a new (non-Ramsey) $\alpha$-tree cover with $3$ trees.
Together with~\Cref{thm:bfn}, this yields the aforementioned~\Cref{cor:3-cover}, giving a tight bound (up to $\log^{3/4}n$) on the distortion of tree covers with $3$ trees. 

The key idea is exploiting the ``redundancy'' of Ramsey tree covers: If two points $x,y \in X$ have \emph{different home trees}, then they have distortion $\alpha$ in two trees among $T_1, T_2, T_3, T_4$, while we only need one such tree.
However, we cannot just naively remove any one tree $T_i$, because then pairs of points $x,y \in X$ \emph{with the same home-tree} $T_i$ would not have a tree with distortion $\alpha$.

Intuitively, the trick to overcome this ``entanglement'' between all four trees is to ``mix them together'', as follows.
Let $F$ be the forest obtained by the disjoint union $T_1 \sq T_2 \sq T_3 \sq T_4$, meaning each $T_i$ is a distinct connected component of $F$.
The new cover $F_1, F_2, F_3$ will consist of three copies of $F$, but \emph{with different embeddings} of $X$ into each of the copies.
(We can turn $F$ into a tree by connecting the $T_i$'s with edges of weight $\infty$, so from now on we allow ourselves to work with ``forest covers''.)

Note that in $F$, there are four natural candidates for embedding a given point $x \in X$, namely, the nodes $T_1 (x), T_2 (x), T_3(x), T_4 (x)$.
Indeed, in each copy $F_j$, we will embed $x$ as one of them.
Formally, for each $F_j$, we will have a corresponding \emph{labeling function} $f_j : X \to [1..4]$; if $x \in X$ has label $\ell$ in $f_j$ (namely $f_j (x) = \ell$), then $F_j$ embeds $x$ in the natural candidate from $\ell$'s tree, that is, $F_j (x) = T_\ell (x)$.

We now define the labelings $f_1,f_2,f_3$. The labels $f_1 (x), f_2(x), f_3(x)$ of a point $x \in X$ will depend only on $\home(x)$. Namely, all points with the same home tree will get the same labels.
The following table defines $f_1, f_2, f_3$:
\[
\begin{array}{c@{\qquad}ccc}
\toprule
\home(x) & f_1(x) & f_2(x) & f_3(x) \\
\midrule
1 & 1 & 4 & 3\\
2 & 1 & 2 & 2\\
3 & 3 & 2 & 3\\
4 & 3 & 4 & 2\\
\bottomrule
\end{array}
\]
The point is that every two points $x,y \in X$ have some ``good'' labeling $f_j$ in which
\begin{itemize}
    \item $x$ and $y$ get the same label $f_j (x) = f_j (y) = \ell$, and
    \item $\ell = \home(x)$ or $\ell = \home(y)$, i.e., the label $\ell$ corresponds to the home tree of  $x$ or of $y$.
\end{itemize}
Therefore, in $F_j$, the points $x$ and $y$ are embedded into $T_\ell (x)$ and $T_\ell (y)$ respectively, and hence
\[
d_{F_j} (x,y) = d_{T_\ell} (x,y) \leq \alpha \cdot d_X (x,y),
\]
where the last inequality is because $T_\ell$ is the home tree of $x$ or of $y$.
Thus, every $x,y \in X$ have some $F_j$ with distortion $\alpha$.
The following table specifies a good labeling $f_j$ for every pair of points $x,y \in X$ according to their home trees:

\[
\begin{tikzpicture}[x=1cm, y=1cm, line width=0.4pt]
  \draw (0,0) -- (7,0);
  \draw (3.4,1.45) -- (3.4,-2.4);
  \draw (0,1.45) -- (3.4,0);

  \node[anchor=south west, inner sep=0pt] at (0.16,0.15)
    {$\home(x)$};
  \node[anchor=north east, inner sep=0pt] at (3.24,1.30)
    {$\home(y)$};

  \foreach \i in {1,...,4} {
    \node at ({3.4+0.9*(\i-0.5)},0.725) {$\i$};
    \node at (1.7,{-0.6*(\i-0.5)}) {$\i$};
  }

  \foreach \row/\col/\idx in {
    1/1/1,
    2/1/1, 2/2/2,
    3/1/3, 3/2/2, 3/3/1,
    4/1/2, 4/2/3, 4/3/1, 4/4/2
  } {
    \node at ({3.4+0.9*(\col-0.5)},{-0.6*(\row-0.5)})
      {$f_{\idx}$};
  }
\end{tikzpicture}
\]
Hence, by starting with the Ramsey tree cover of~\Cref{thm:bfn} with $k = 4$ trees, we obtain Corollary~\ref{cor:3-cover}.

Let us now discuss the limitation of the above basic approach, for a general number of trees $k \geq 1$ in the $\alpha$-Ramsey cover.
We argue that this trick alone cannot yield an $\alpha$-tree cover with fewer than $\Omega(k)$ trees (i.e., it only allows a constant factor decrease in the number of trees).
To see this, fix a labeling $f$ in which $f(x)$ depends only on $x$'s home tree.
Suppose $f$ is good for a pair $x,y \in X$.
Then $f(x)=\home(y)$ or $f(y)=\home(x)$, and in either case, the pair $\{\home(x), \home(y)\}$ is of the form $\{\home(z),f(z)\}$ for $z\in\{x,y\}$.
As $\home(z)$ can have at most $k$ different values, this means that $f$ can be good for at most $k$ pairs of home trees.
So, in order to take care of all $\binom{k}{2} + k = k(k+1)/2$ such pairs, we have to use at least $(k+1)/2$ labeling functions.

Therefore, to prove the main result of~\Cref{thm:main}, we have to introduce some additional ideas, which importantly allow the labels to not depend solely on the home tree of the point by incurring an additional constant factor in distortion.

\section{Proof of Main~\Cref{thm:main}}\label{sec:proof-of-main}

We will use the additional notion of \emph{pivots}, as follows.
For a point $x \in X$ and $1 \leq i \leq k$, the $i$-pivot of $x$ is a point $p_i (x)$ with home tree $T_i$ which is closest to $x$ (ties broken arbitrarily).
That is, $d_X (x,p_i (x)) = \min\{d_X (x,z) \mid z \in H_i\}$.
Intuitively, instead of insisting on a common labeling for a pair $x,y$ which sends them to one of their homes, we settle for a labeling that sends them to the home of a ``close enough'' pivot.
As in the warm-up, we let $F$ be the forest obtained by the disjoint union $T_1 \sq T_2 \sq \cdots \sq T_k$, and the new cover $F_1, F_2, \dots $ consists of $O(\sqrt{k})$ copies of $F$ but with different embeddings of $X$ into each of them.
Again, the embedding into $F_j$ will be based on a corresponding labeling function $f_j : X \to [1..k]$ such that $F_j (x) = T_\ell (x)$ for $\ell = f_j (x)$.
We use the pivots to relax the ``good labeling'' property of the collection $\{f_j\}_j$ from the warm-up.
Formally, we will construct the labelings $\{f_j \}_j$ to have the following property for every pair of points $x,y \in X$: 
\begin{enumerate}[label=\textbf{(P)}]         
    \item There exists some labeling $f_j$ where $x$ and $y$ get the same label $f_j (x) = f_j (y) = \ell$, such that one of $x,y$ has distance at most $d_X (x,y)$ to its $\ell$-pivot.\label{prop:P}
\end{enumerate}

The following claim shows that ensuring Property~\ref{prop:P} yields a tree cover with distortion $3\alpha$.
\begin{claim}\label{claim:pivot-implies-cover}
If Property~\ref{prop:P} holds for every $x,y \in X$, then $F_1,F_2, \dots$ is a $3\alpha$-tree cover of $X$.
\end{claim}
\begin{proof}
Fix $x,y \in X$ and let $f_j$ be a labeling for which $f_j (x) = f_j (y) = \ell$ and one of $x,y$ has distance at most $d_X (x,y)$ to its $\ell$-pivot, say $d_X(x, p_\ell(x)) \leq d_X(x,y)$, guaranteed by Property~\ref{prop:P}.
Then,
\begin{align*}
    d_{F_j} (x,y) &= d_{T_\ell} (x,y) &&\text{as $f_j (x) = f_j (y) = \ell$,} \\
    &\leq d_{T_\ell}(x,p_\ell (x)) + d_{T_\ell} (p_\ell (x), y) && \text{by triangle inequality,} \\
    &\leq \alpha \cdot \big(d_X (x,p_\ell (x)) + d_X (p_\ell(x), y) \big) && \text{as $T_\ell$ is the home tree of $p_\ell (x)$,} \\
    &\leq \alpha \cdot \big(d_X (x,y) + d_X (p_\ell (x), x) + d_X (x,y) \big) && \text{by triangle inequality,} \\
    &\leq 3\alpha \cdot d_X (x,y) && \text{as $x$ is closer to $p_\ell (x)$ than to $y$,}
\end{align*}
as needed.
\end{proof}

So, from now on, we focus on the following \emph{labeling problem}: Construct a set of $O(\sqrt{k})$ labelings $\{f_j : X \to [1..k] \}_j$ which satisfies Property~\ref{prop:P} for every pair of points $x,y \in X$.

\subsection{The bipartite labeling problem}
The central part of the construction is based on a certain ``bipartite variant'' of our problem. In this variant, $X$ is partitioned into two sides as $X = A \sq B$, and the goal is to construct a set of $O(\sqrt{k})$ labelings that satisfies Property~\ref{prop:P} only for the pairs $x,y$ with $x \in A$ and $y \in B$.
The simple reduction from the general setting to the bipartite one is based on recursively partitioning the points into two sides according to their home trees, and it is discussed in detail later, in Section~\ref{sec:reduction-to-bipartite}.
In this section, we show:
\begin{claim}\label{lem:bipartite-labeling}
    The bipartite labeling problem can be solved with $2 \lceil\sqrt{k}\rceil$ labeling functions.
\end{claim}

\begin{proof}
The bipartiteness is used for symmetry breaking in Property~\ref{prop:P}: For $x \in A$ and $y \in B$, we will ensure a common labeling $f_j (x) = f_j (y) = \ell$ such that $d_X (x,p_\ell (x)) \leq d_X (x,y)$.
Thus, points in $A$ and points in $B$ will play different roles in the construction.

The general setting is as follows.
We arbitrarily arrange the trees $T_1, \dots, T_k$ in a $\lceil \sqrt{k} \rceil \times \lceil \sqrt{k} \rceil$ matrix $M$.
(Note that the matrix has $(\lceil \sqrt{k} \rceil)^2 - k$ empty entries, and this will not be a problem.)
We will have two types of labeling functions, for rows and for columns:
\begin{itemize}
    \item The labeling $r_i : X \to [1..k]$  is for the $i$-th row of $M$.
    \item The labeling $c_j : X \to [1..k]$ is for the $j$-th column of $M$.
\end{itemize}
So in total, we have exactly $2 \lceil \sqrt{k} \rceil$ labelings, as needed.

\begin{description}
    \item[Labels of $A$:] 
We start by defining these labelings on the set $A$.
This is done by executing a separate process for each $x \in A$, which determines its labels $r_i (x)$ and $c_j (x)$ for every $i$ and $j$.
In this process, $x$ starts from the full matrix $M$. It iteratively ``deletes'' entries and decides on some of the labels, until every entry is deleted.
An iteration works as follows:
\begin{itemize}
    \item Find a non-deleted entry $(i,j)$ such that the distance from $x$ to its pivot in the tree $T_\ell$ found in this entry is minimized. 
    \item Assign $\ell$ as $x$'s labels for the $i$-row and $j$-column, i.e., $r_i (x) := \ell$ and $c_j (x) := \ell$.
    
    \item Delete all entries in the $i$-th row, and all entries in the $j$-th column.
\end{itemize}
Note that the deletions guarantee that each row/column label of $x$ is assigned at most once during this process.
After the process halts and the matrix is all deleted, there might still be some undefined row/column labels of $x$ (because the matrix might have some initially empty entries).
These labels are just set arbitrarily.

\item[Labels of $B$:]
Next, we define the labelings on the set $B$.
This process is simpler.
Given $y \in B$, we find the entry $(i_y,j_y)$ containing its home tree $T_h$ for $h = \home(y)$.%
\begin{itemize}
    \item For each row $i$ such that entry $(i,j_y)$ contains some tree $T_\ell$, we set $r_i (y) := \ell$.

    \item For each column $j$ such that entry $(i_y,j)$ contains some tree $T_\ell$, we set $c_j (y) := \ell$. 

    \item Any other row/column label of $y$ is set arbitrarily.
\end{itemize}
\end{description}

The crux of these two different strategies for $A$ and for $B$ is that they nicely \emph{correlate} with each other.
This is expressed in the following argument showing that these labelings indeed solve our bipartite problem; see Figure~\ref{fig:bipartite-labeling} for an illustration.
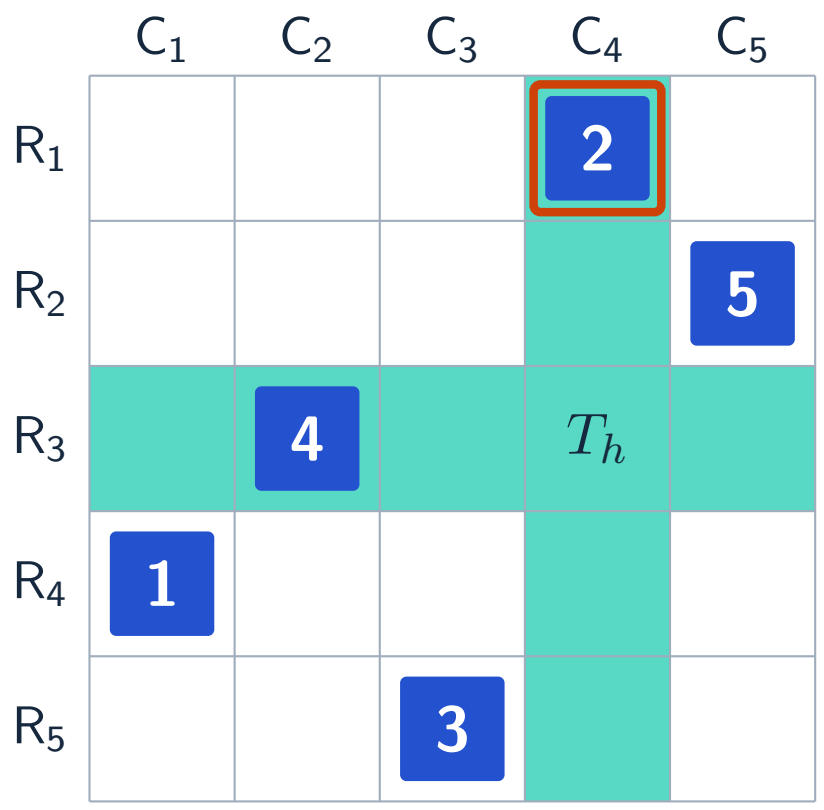
\begin{figure}[t]
\centering
{\fontfamily{lmss}\selectfont
\definecolor{smInk}{HTML}{16283E}%
\definecolor{smBlue}{HTML}{2452CF}%
\definecolor{smTeal}{HTML}{57D9C5}%
\definecolor{smOrange}{HTML}{CE4108}%
\definecolor{smGrid}{HTML}{A0ADBD}%
\begin{tikzpicture}[x=.96cm,y=-.96cm,text=smInk,
  axis label/.style={font=\small,inner sep=0pt},
  agreement/.style={draw=smOrange,line width=1.6pt,rounded corners=1.3pt}]
  \def\homeRow{3}
  \def\homeCol{4}

  \fill[smTeal] (0,\homeRow-1) rectangle (5,\homeRow);
  \fill[smTeal] (\homeCol-1,0) rectangle (\homeCol,5);

  \foreach \i in {0,...,5}{
    \draw[smGrid,line width=.4pt] (0,\i) -- (5,\i);
    \draw[smGrid,line width=.4pt] (\i,0) -- (\i,5);
  }
  \foreach \i in {1,...,5}{
    \node[axis label] at (\i-.5,-.25) {C\textsubscript{\i}};
    \node[axis label] at (-.34,\i-.5) {R\textsubscript{\i}};
  }
  \node[inner sep=0pt] at (\homeCol-.5,\homeRow-.5)
    {$T_h$};

  \foreach \r/\c/\s in {4/1/1,1/4/2,5/3/3,3/2/4,2/5/5}{
    \fill[smBlue,rounded corners=1.2pt]
      (\c-.86,\r-.86) rectangle (\c-.14,\r-.14);
    \node[font=\large\bfseries,text=white] at (\c-.5,\r-.5) {\s};
    \ifnum\s=2\relax
      \draw[agreement] (\c-.94,\r-.94) rectangle (\c-.06,\r-.06);
    \fi
  }
\end{tikzpicture}}
\caption{Given $(x,y) \in A \times B$, the numbered cells show the choices of $x\in A$ in order, and the shaded cross marks the row and column of $T_h$ for $h = \home(y)$. 
The highlighted cell contains the tree chosen by $x$ in the iteration that deleted $T_h$ from the matrix.}
\label{fig:bipartite-labeling}
\end{figure}

Fix some pair $(x,y) \in A \times B$.
Consider the iteration in which $x$ deleted the entry $(i_y, j_y)$ containing $T_h$ for $h = \home(y)$.
Let $(i,j)$ be the entry found by $x$ in this iteration, and let $T_\ell$ be the tree in this entry, so $r_i (x) = c_j (x) = \ell$.
Since $(i,j)$ was chosen and $(i_y, j_y)$ was a potential choice, we have
\begin{equation}\label{eq:closer-to-pivot}
    d_X (x,p_\ell (x)) \leq d_X (x, p_h (x)) \leq d_X (x,y) 
\end{equation}
(the second inequality is since $y \in H_h$ as $h = \home(y)$).
As $x$ deletes the $i$-th row and the $j$-column in this iteration, we have that either $i=i_y$ or $j = j_y$.
In the former case $c_j (y) = \ell$, and in the latter, $r_i (y) = \ell$.
Thus, we have shown that one of the labelings $r_i$ or $c_j$ assigns both $x$ and $y$ the label $\ell$.
Together with~\Cref{eq:closer-to-pivot}, this establishes Property~\ref{prop:P} for $x,y$, which concludes the proof.
\end{proof}

\subsection{Reduction to the bipartite case}\label{sec:reduction-to-bipartite}

We now give the reduction from the general labeling problem to the bipartite one, showing that:

\begin{claim}\label{claim:reduction-to-bipartite}
    The general labeling problem can be solved with $O(\sqrt{k})$ labeling functions:
    Namely, there are $O(\sqrt{k})$ labeling functions $\{f_j: X \to [1..k]\}_j$ that ensure Property~\ref{prop:P} for every pair $x,y \in X$.
\end{claim}

As discussed at the beginning of Section~\ref{sec:proof-of-main}, the above Claim~\ref{claim:reduction-to-bipartite} together with Claim~\ref{claim:pivot-implies-cover} imply the main Theorem~\ref{thm:main}. 

\begin{proof}[Proof of Claim~\ref{claim:reduction-to-bipartite}.]
    We first show that when $k$ is a power of $2$, the general labeling problem can be solved with at most $15 \lceil \sqrt{k} \rceil$ labeling functions, by induction on $k$.
    The base cases are when $k \leq 15^2$.
    Then, $15 \lceil\sqrt{k}\rceil \geq k$, so we can use the trivial collection of $k$ labeling functions given by the Ramsey tree cover itself, namely, $f_1 \equiv 1, f_2 \equiv 2, \dots, f_k \equiv k$, which clearly satisfies Property~\ref{prop:P} for every pair $x,y \in X$ (say, with $f_\ell$ such that $\ell = \home(x)$).
    
    The induction step works as follows.
    Partition $X = A \sq B$ where $A := H_1 \sq \cdots \sq H_{k/2}$ and $B := H_{k/2+1} \sq \cdots \sq H_k$.
    As $T_1, \dots, T_{k/2}$ is an $\alpha$-Ramsey tree cover for $(A,d_X)$, by the induction hypothesis we can find $15 \lceil\sqrt{k/2}\rceil$ labeling functions $\{f^A_j : A \to [1..k/2]\}_j$ such that Property~\ref{prop:P} holds for every $x,y \in A$.
    Similarly, we find $15\lceil\sqrt{k/2}\rceil$ labeling functions $\{f^B_j : B \to [k/2+1..k]\}_j$ such that Property~\ref{prop:P} holds for every $x,y \in B$.
    We concatenate each $f^A_j$ and $f^B_j$ into $f_j : X \to [1..k]$ in the obvious way:
    \[
        f_j (x) = 
        \begin{cases}
            f_j^A (x) & \text{if $x \in A$,} \\
            f_j^B (x) & \text{if $x \in B$.}
        \end{cases}
    \]
    Then $\{f_j : X \to [1..k]\}_j$ satisfies Property~\ref{prop:P} for every pair of points $x,y$ which are both from $A$ or both from $B$.
    To handle the pairs with $x \in A$ and $y \in B$, we apply Claim~\ref{lem:bipartite-labeling} and get an additional set of $2 \lceil\sqrt{k}\rceil$ labeling functions that ensure Property~\ref{prop:P} for them.
    We add these labeling functions to our collection, thus ensuring Property~\ref{prop:P} for every pair of points $x,y \in X$.
    Overall, the number of labeling functions defined in the induction step is
    \[
    15 \left\lceil\sqrt{\tfrac{k}{2}}\right\rceil + 2 \lceil \sqrt{k}\rceil \leq 15 \big( \sqrt{\tfrac{k}{2}} + 1 \big) + 2 \lceil \sqrt{k}\rceil
    \leq \big(\underbrace{\tfrac{15}{\sqrt{2}} + 1 + 2}_{\approx 13.607} \big)\lceil \sqrt{k} \rceil \leq 15 \lceil\sqrt{k}\rceil
    \]
    (where the penultimate inequality used $15 < \sqrt{k}$), as needed.

    Finally, when $k$ is not a power of $2$, we can duplicate trees in the Ramsey tree cover to get to the smallest power of $2$ larger than $k$, which is at most $2k$, and apply the above to get at most $15 \lceil \sqrt{2k} \rceil = O(\sqrt{k})$ labeling functions.
\end{proof}

\paragraph{Acknowledgments.}
E.T. thanks Merav Parter for encouraging  discussions that led to this work.

\bibliographystyle{alphaurl}
\bibliography{ref}

\end{document}